\documentstyle[prb,aps]{revtex}
\begin{document}
\draft
                                                            
\title{Exact Dynamical Structure Factor of a Bose Liquid}

\author{Girish S. Setlur}
\address{Department of Physics and Materials Research Laboratory
\\ University of Illinois at Urbana-Champaign, Urbana, Illinois 61801}

\date{\today}
\maketitle

                                                  
\begin{abstract}

 Based on ideas introduced in a previous preprint cond-mat/9701206 we propose
 an exactly solvable model of bosons interacting amongst themselves
 via a Van-der Waal-like repulsive interaction, and compute both the
 filling fraction and the dynamical structure factor
 exactly. The novelty of this approach involves introducing, analogous to
 Fermi sea (or surface) displacements,
 Bose fields that in this case, correspond
 to fluctuations of the Bose condensate. The exact dynamical structure factor
 has a coherent part that corresponds to the Bogoliubov spectrum and an
 incoherent part that is a result of correlations.

\end{abstract}

\section{Introduction}
 
 The microscopic thoery of He4 is by now a well established and extremely
 succesful theory. This theory is the culmination of the efforts of a
 large number of distinguished physicists starting from Bogoliubov
 \cite{Bogoliubov} whose theory of a weakly non-ideal bose gas was the
 inspiration for all the theoretical work that followed in this field. 
 However, the modern theory of superfluidity as it is understood today
 was poineered by Penrose and Onsager \cite{Penrose} who introduced the
 concept of off-diagonal long range order.  First attempts
 at guessing the form of the excitation
 spectrum was made by Bijl and Feynman \cite{Bijl}. Landau \cite{Landau}
 surmised many of the properties of the excitation spectrum including the
 phonon and the roton parts before it was measured experimaentally. 
 The microscopic theory
 for obtaining the static structure factor employs correlated basis
 functions that was pioneered by Bijl, Dingje and Jastrow \cite{Jastrow}.
 An elaborate application of this technique to quantum liquids in
 general may be found in the book by Feenberg \cite{Feenberg}. 
 Also the work of Kebukawa, Sunakawa and Yamasaki \cite{Sun}
 in computing the dynamical
 properties including the excitation spectrum, backflow e.t.c. has to be
 mentioned.

	In the light of these historical remarks, it seems that yet another
 theory of this system is redundant. However, there is always room for
 a fresh perspective on well understood problems. The hope is that
 the approach that we are now going to describe along with its fermionic
 counterpart (described elsewhere \cite{Setlur1} )
 will produce a model of matter that has these desirable qualities,
 namely, it is exactly solvable, captures the important physics and
 contains no phenomenological parameters other than the electric
 charge and mass. If thse types of models were to
 prove succesful, an outcome that is tantalisingly plausible, 
 it will be a vindication of the soundness of the reductionist philosophy.
 
\section{ Bosons at zero temperature }

 In this article we do what we did earlier (cond-mat/9701206)
 except that here we
 are bosonizing the bosons. Let $ b_{ {\bf{k}} } $
 and  $ b^{\dagger}_{ {\bf{k}} } $ be the Bose fields in question.
 Analogous to Fermi sea displacements we introduce Bose
 condensate displacements ( $ d_{ {\bf{k}} }({\bf{q}}) $) which are also
 bose fields obeying canonical commutationrelations identical to its
 parents $  b_{ {\bf{k}} } $ and $  b^{\dagger}_{ {\bf{k}} } $.
\[
b^{\dagger}_{ {\bf{k+q/2}} } b_{ {\bf{k-q/2}} }
 = N \mbox{  }\delta_{ {\bf{k}}, 0 } \delta_{ {\bf{q}}, 0 }
 + (\sqrt{N})
[\delta_{ {\bf{k+q/2}}, 0 }\mbox{  }d_{ {\bf{k}} }(-{\bf{q}})
 + \delta_{ {\bf{k-q/2}}, 0 }\mbox{  }d^{\dagger}_{ {\bf{k}} }({\bf{q}})]
\]
\[
 + \sum_{ {\bf{q}}_{1} }T_{1}({\bf{k}}, {\bf{q}}, {\bf{q}}_{1})
 d^{\dagger}_{ {\bf{k+q/2-q_{1}/2}} }({\bf{q}}_{1})
 d_{ {\bf{k-q_{1}/2}} }(-{\bf{q}}+{\bf{q}}_{1})
\]
\begin{equation}
 +  \sum_{ {\bf{q}}_{1} }T_{2}({\bf{k}}, {\bf{q}}, {\bf{q}}_{1})
 d^{\dagger}_{ {\bf{k-q/2+q_{1}/2}} }({\bf{q}}_{1})
 d_{ {\bf{k+q_{1}/2}} }(-{\bf{q}}+{\bf{q}}_{1})
\end{equation}
 The above relation is meant to be an operator identity. In other words, all
 dynamical moments of the product
 $ b^{\dagger}_{ {\bf{k+q/2}} } b_{ {\bf{k-q/2}} } $ are equal when evaluated
 either in the original bose language or in the condensate-displacement
 language.  This is true provided we identify the vacuum of the 
 $ d_{ {\bf{k}} }({\bf{q}}) $ with the bose condensate (the ground
 state of the noninteracting system).
\begin{equation}
d_{ {\bf{k}} }({\bf{q}})|BC\rangle = 0 
\end{equation}
 Furthermore, the chemical potential of the condesate-displacement bosons
 is taken to be zero.  In other words the number of condensate displacement 
 bosons is not consereved.

To make all the dynamical moments of
 $ b^{\dagger}_{ {\bf{k+q/2}} } b_{ {\bf{k-q/2}} } $ come out right and the
 commutation rules amongst them also come out right provided we choose
 $ T_{1} $ and $ T_{2} $ such that,
\[
(\sqrt{N})^{2}\delta_{ {\bf{k+q/2}}, 0 }
\delta_{ {\bf{k^{''}-q^{''}/2}}, 0 }
\sum_{ {\bf{q}}_{1} }T_{1}({\bf{k}}^{'}, {\bf{q}}^{'}, {\bf{q}}_{1})
\langle d_{ {\bf{k}} }(-{\bf{q}})
d^{\dagger}_{ {\bf{k}}^{'} + {\bf{q}}^{'}/2 - {\bf{q}}_{1}/2 }
({\bf{q}}_{1})d_{ {\bf{k}}^{'} - {\bf{q}}_{1}/2 }
(-{\bf{q}}^{'} + {\bf{q}}_{1}) d^{\dagger}_{ {\bf{k}}^{''} }({\bf{q}}^{''})
 \rangle
\]
\[
+(\sqrt{N})^{2}\delta_{ {\bf{k+q/2}}, 0 }
\delta_{ {\bf{k^{''}-q^{''}/2}}, 0 }
\sum_{ {\bf{q}}_{1} }T_{2}({\bf{k}}^{'}, {\bf{q}}^{'}, {\bf{q}}_{1})
\langle d_{ {\bf{k}} }(-{\bf{q}})
d^{\dagger}_{ {\bf{k}}^{'} - {\bf{q}}^{'}/2 + {\bf{q}}_{1}/2 }
({\bf{q}}_{1})d_{ {\bf{k}}^{'} + {\bf{q}}_{1}/2 }
(-{\bf{q}}^{'} + {\bf{q}}_{1}) d^{\dagger}_{ {\bf{k}}^{''} }({\bf{q}}^{''})
 \rangle
\]
 =
\begin{equation}
N\mbox{ }\delta_{ {\bf{k+q/2}}, 0 }
\delta_{ {\bf{k+q/2}}, {\bf{k^{''}-q^{''}/2}} }
\delta_{ {\bf{k-q/2}}, {\bf{k^{'}+q^{'}/2}} }
\delta_{ {\bf{k^{'}-q^{'}/2}}, {\bf{k^{''}+q^{''}/2}} }
\end{equation}This means,
\begin{equation}
T_{1}( -{\bf{q}}-{\bf{q}}^{'}/2, {\bf{q}}^{'}, -{\bf{q}} )
 = 1; \mbox{   }{\bf{q}} \neq 0; \mbox{   }{\bf{q}}^{'} \neq 0
\end{equation}
\begin{equation}
T_{2}( -{\bf{q}}^{'}/2, {\bf{q}}^{'}, -{\bf{q}} )
 = 0; \mbox{   }{\bf{q}} \neq 0; \mbox{   }{\bf{q}}^{'} \neq 0
\end{equation}
in order for the kinetic energy operator to have the form,
\begin{equation}
K = \sum_{ {\bf{k}} }\epsilon_{ {\bf{k}} } \mbox{  }
d^{\dagger}_{ (1/2){\bf{k}} }({\bf{k}})
\mbox{ }d_{ (1/2){\bf{k}} }({\bf{k}})
\end{equation}
we must have,
\begin{equation}
\epsilon_{ {\bf{k+q_{1}/2}} }T_{1}( {\bf{k+q_{1}/2}}, {\bf{0}}, {\bf{q}}_{1})
 + \epsilon_{ {\bf{k-q_{1}/2}} }T_{2}( {\bf{k-q_{1}/2}}, {\bf{0}}, {\bf{q}}_{1})
 = \delta_{ {\bf{k}}, {\bf{q}}_{1}/2 }\epsilon_{ {\bf{q}}_{1} }
\label{EQ1}
\end{equation}
In order for,
\[
  \sum_{ {\bf{k}} }b^{ \dagger }_{ {\bf{k}} }b_{ {\bf{k}} }
 = N
\]
we must have,
\begin{equation}
T_{1}({\bf{k+q_{1}/2}}, {\bf{0}}, {\bf{q_{1}}})
 + T_{2}({\bf{k-q_{1}/2}}, {\bf{0}}, {\bf{q_{1}}})
 = 0
\label{EQ2}
\end{equation}
 In order for the commutation rules amongst the
 $ b^{\dagger}_{ {\bf{k+q/2}} } b_{ {\bf{k-q/2}} } $ to come out right,
 we must have in addition to all the above relations,
\begin{equation}
T_{2}({\bf{q}}/2, {\bf{q}}, -{\bf{q}}^{'}) = 0
\end{equation}
\begin{equation}
T_{2}(-{\bf{q}}/2, {\bf{q}}, {\bf{q}}+{\bf{q}}^{'}) = 0
\end{equation}
Any choice of $ T_{1} $ and $ T_{2} $ consistent with the above
 relations should suffice.
From the above relations (Eq.(~\ref{EQ1}) and Eq.(~\ref{EQ2})),
 we have quite unambiguously,
\begin{equation}
T_{1}({\bf{k+q_{1}/2}}, {\bf{0}}, {\bf{q}}_{1}) =
\delta_{ {\bf{k}}, {\bf{q}}_{1}/2 }
\end{equation}
\begin{equation}
T_{2}({\bf{k-q_{1}/2}}, {\bf{0}}, {\bf{q}}_{1}) =
-\delta_{ {\bf{k}}, {\bf{q}}_{1}/2 }
\end{equation}
\begin{equation}
T_{1}(-{\bf{q}}-{\bf{q}}^{'}/2, {\bf{q}}^{'}, -{\bf{q}}) = 1
\end{equation}
and,
\begin{equation}
T_{1}({\bf{k}}, {\bf{q}}, {\bf{q}}_{1}) = 0; \mbox{  }otherwise
\end{equation}
\begin{equation}
T_{2}({\bf{k}}, {\bf{q}}, {\bf{q}}_{1}) = 0; \mbox{  }otherwise
\end{equation}
This means that we may rewrite the formula for
 $ b^{\dagger}_{ {\bf{k+q/2}} }b_{ {\bf{k-q/2}} } $
as follows,
\[
b^{\dagger}_{ {\bf{k+q/2}} }b_{ {\bf{k-q/2}} }
 = N \delta_{ {\bf{k}}, 0 }\delta_{ {\bf{q}}, 0 } +
(\sqrt{N})(1-\delta_{ {\bf{k}}, 0 }\delta_{ {\bf{q}}, 0 })
[\delta_{ {\bf{k+q/2}}, 0 }d_{ {\bf{k}} }(-{\bf{q}})
 + \delta_{ {\bf{k-q/2}}, 0 }d^{\dagger}_{ {\bf{k}} }({\bf{q}})]
\]
\[
+ 
d^{\dagger}_{ (1/2){\bf{k+q/2}} }({\bf{k+q/2}})
d_{ (1/2){\bf{k-q/2}} }({\bf{k-q/2}})
\]
\begin{equation}
-\delta_{ {\bf{k}}, 0 }\delta_{ {\bf{q}}, 0 }
\sum_{ {\bf{q}}_{1} }d^{\dagger}_{ {\bf{q}}_{1}/2 }({\bf{q}}_{1})
d_{ {\bf{q}}_{1}/2 }({\bf{q}}_{1})
\end{equation}
Consider an interaction of the type,
\begin{equation}
H_{I} = (\frac{\rho_{0}}{2})
\sum_{ {\bf{q}} \neq 0 }
v_{ {\bf{q}} }\sum_{ {\bf{k}}, {\bf{k}}^{'} }
[\delta_{ {\bf{k+q/2}}, 0 }\mbox{  }d_{ {\bf{k}} }(-{\bf{q}})
 + \delta_{ {\bf{k-q/2}}, 0 }\mbox{  }d^{\dagger}_{ {\bf{k}} }({\bf{q}})]
[\delta_{ {\bf{k^{'}-q/2}}, 0 }\mbox{  }d_{ {\bf{k}}^{'} }({\bf{q}})
 + \delta_{ {\bf{k^{'}+q/2}}, 0 }\mbox{  }
d^{\dagger}_{ {\bf{k}}^{'} }(-{\bf{q}})]
\end{equation}
or,
\begin{equation}
H_{I} = (\frac{\rho_{0}}{2})
\sum_{ {\bf{q}} \neq 0 }
v_{ {\bf{q}} }
[d_{ -{\bf{q}}/2 }(-{\bf{q}})
 + d^{\dagger}_{ {\bf{q}}/2 }({\bf{q}})]
[d_{ {\bf{q}}/2 }({\bf{q}})
 + d^{\dagger}_{ -{\bf{q}}/2 }(-{\bf{q}})]
\end{equation}
and the free case is given by,
\begin{equation}
H_{0} = \sum_{ {\bf{q}} }\epsilon_{ {\bf{q}} }
d^{\dagger}_{ (1/2){\bf{q}} }({\bf{q}})d_{ (1/2){\bf{q}} }({\bf{q}})
\end{equation}
The full hamiltonian may be diagonalised as follows,
\begin{equation}
H = \sum_{ {\bf{q}} }\omega_{ {\bf{q}} }
f^{\dagger}_{ {\bf{q}} }f_{ {\bf{q}} }
\end{equation}
and,
\begin{equation}
f_{ {\bf{q}} } =
 (\frac{ \omega_{ {\bf{q}} } + \epsilon_{ {\bf{q}} }
 + \rho_{0}v_{ {\bf{q}} } }
{ 2 \mbox{ }\omega_{ {\bf{q}} } })^{\frac{1}{2}}
d_{ {\bf{q}}/2 }({\bf{q}})
 + (\frac{ -\omega_{ {\bf{q}} } +
 \epsilon_{ {\bf{q}} } + \rho_{0}v_{ {\bf{q}} } }
{ 2 \mbox{ }\omega_{ {\bf{q}} } })^{\frac{1}{2}}
d^{\dagger}_{ -{\bf{q}}/2 }(-{\bf{q}})
\end{equation}
\begin{equation}
f^{\dagger}_{ -{\bf{q}} } =
(\frac{ -\omega_{ {\bf{q}} } +
 \epsilon_{ {\bf{q}} } + \rho_{0}v_{ {\bf{q}} } }
{ 2 \mbox{ }\omega_{ {\bf{q}} } })^{\frac{1}{2}}
d_{ {\bf{q}}/2 }({\bf{q}})
+
 (\frac{ \omega_{ {\bf{q}} } + \epsilon_{ {\bf{q}} }
 + \rho_{0}v_{ {\bf{q}} } }
{ 2 \mbox{ }\omega_{ {\bf{q}} } })^{\frac{1}{2}}
d^{\dagger}_{ -{\bf{q}}/2 }(-{\bf{q}})
\end{equation}
\begin{equation}
d_{ {\bf{q}}/2 }({\bf{q}})
 = (\frac{ \omega_{ {\bf{q}} } + \epsilon_{ {\bf{q}} }
 + \rho_{0}v_{ {\bf{q}} } }
{ 2 \mbox{ }\omega_{ {\bf{q}} } })^{\frac{1}{2}}f_{ {\bf{q}} }
 - (\frac{ -\omega_{ {\bf{q}} } + \epsilon_{ {\bf{q}} }
 + \rho_{0}v_{ {\bf{q}} } }
{ 2 \mbox{ }\omega_{ {\bf{q}} } })^{\frac{1}{2}}f^{\dagger}_{ -{\bf{q}} }
\end{equation}
\begin{equation}
d^{\dagger}_{ -{\bf{q}}/2 }(-{\bf{q}})
 = (\frac{ \omega_{ {\bf{q}} } + \epsilon_{ {\bf{q}} }
 + \rho_{0}v_{ {\bf{q}} } }
{ 2 \mbox{ }\omega_{ {\bf{q}} } })^{\frac{1}{2}}f^{\dagger}_{ -{\bf{q}} }
 - (\frac{ -\omega_{ {\bf{q}} } + \epsilon_{ {\bf{q}} }
 + \rho_{0}v_{ {\bf{q}} } }
{ 2 \mbox{ }\omega_{ {\bf{q}} } })^{\frac{1}{2}}f_{ {\bf{q}} }
\end{equation}
\begin{equation}
\omega_{ {\bf{q}} } = \sqrt{ \epsilon^{2}_{ {\bf{q}} }
+ 2 \rho_{0}v_{ {\bf{q}} }\epsilon_{ {\bf{q}} } }
\end{equation}
                    
From this we may deduce,
\begin{equation}
\langle d^{\dagger}_{ (1/2){\bf{q}} }({\bf{q}})
d_{ (1/2){\bf{q}} }({\bf{q}}) \rangle
 = \frac{ -\omega_{ {\bf{q}} } + \epsilon_{ {\bf{q}} }
+ \rho_{0}v_{ {\bf{q}} } }{ 2\mbox{ }\omega_{ {\bf{q}} } }
\end{equation}
 It is worth while to note that 
 $ \epsilon_{ {\bf{q}} } + \rho_{0}v_{ {\bf{q}} } > \omega_{ {\bf{q}} } $.
From this it is possible to write down the filling fraction.
\newline
{\center{ {\bf{FILLING FRACTION}} }}
\begin{equation}
f_{0} = N_{0}/N = 1 -
(1/N)\sum_{ {\bf{q}} }\langle d^{\dagger}_{ (1/2){\bf{q}} }({\bf{q}})
d_{ (1/2){\bf{q}} }({\bf{q}}) \rangle
\end{equation}
or,
\begin{equation}
f_{0} = N_{0}/N = 1 - (1/2\pi^{2}\rho_{0})\int_{0}^{\infty}
 \mbox{ }dq\mbox{ }
q^{2} ( \frac{ -\omega_{ {\bf{q}} } + \epsilon_{ {\bf{q}} }
+ \rho_{0}v_{ {\bf{q}} } }{ 2\mbox{ }\omega_{ {\bf{q}} } } )
\end{equation}
First define,
\begin{equation}
A_{ {\bf{q}} } = (\frac{ \omega_{ {\bf{q}} } + \epsilon_{ {\bf{q}} }
+ \rho_{0}v_{ {\bf{q}} } }{2 \mbox{ }\omega_{ {\bf{q}} } })^{\frac{1}{2}}
\end{equation}
\begin{equation}
B_{ {\bf{q}} } = (\frac{ -\omega_{ {\bf{q}} } + \epsilon_{ {\bf{q}} }
+ \rho_{0}v_{ {\bf{q}} } }{2 \mbox{ }\omega_{ {\bf{q}} } })^{\frac{1}{2}}
\end{equation}
The density operator is,
\begin{equation}
\rho_{ {\bf{q}} }(t)
  = \sqrt{N}[d_{ -(1/2){\bf{q}} }(-{\bf{q}})(t) +
d^{\dagger}_{ (1/2){\bf{q}} }({\bf{q}})(t)]
+ \sum_{ {\bf{k}} }d^{\dagger}_{ (1/2){\bf{k+q/2}} }({\bf{k+q/2}})(t)
d_{ (1/2){\bf{k-q/2}} }({\bf{k-q/2}})(t)
\end{equation}
\begin{equation}
d_{ -(1/2){\bf{q}} }(-{\bf{q}})(t)
 = A_{ {\bf{q}} }f_{ -{\bf{q}} }e^{-i\mbox{ }\omega_{ {\bf{q}} }t}
 - B_{ {\bf{q}} }f^{\dagger}_{ {\bf{q}} }e^{i\mbox{ }\omega_{ {\bf{q}} }t}
\end{equation}
\begin{equation}
d^{\dagger}_{ (1/2){\bf{q}} }({\bf{q}})(t)
 = A_{ {\bf{q}} }f^{\dagger}_{ {\bf{q}} }e^{i\mbox{ }\omega_{ {\bf{q}} }t}
 - B_{ {\bf{q}} }f_{ -{\bf{q}} }e^{-i\mbox{ }\omega_{ {\bf{q}} }t}
\end{equation}
\begin{equation}
d_{ (1/2){\bf{k-q/2}} }({\bf{k-q/2}})(t)
 = A_{ {\bf{k-q/2}} }f_{ {\bf{k-q/2}} }e^{-i\mbox{ }\omega_{ {\bf{k-q/2}} }t}
 - B_{ {\bf{k-q/2}} }f^{\dagger}_{ {\bf{-k+q/2}} }
e^{i\mbox{ }\omega_{ {\bf{k-q/2}} }t}
\end{equation}
\begin{equation}
d^{\dagger}_{ (1/2){\bf{k+q/2}} }({\bf{k+q/2}})(t)
 = A_{ {\bf{k+q/2}} }f^{\dagger}_{ {\bf{k+q/2}} }
e^{i\mbox{ }\omega_{ {\bf{k+q/2}} }t}
 - B_{ {\bf{k+q/2}} }f_{ {\bf{-k-q/2}} }e^{-i\mbox{ }\omega_{ {\bf{k+q/2}} }t}
\end{equation}
Now define,
\[
S^{>}({\bf{q}}t) = \langle \rho_{ {\bf{q}} }(t) \rho_{ -{\bf{q}} }(0)\rangle
 = N\langle [ d_{ -(1/2){\bf{q}} }(-{\bf{q}})(t)
 + d^{\dagger}_{ (1/2){\bf{q}} }({\bf{q}})(t) ]
[ d_{ (1/2){\bf{q}} }({\bf{q}})(0)
 + d^{\dagger}_{ -(1/2){\bf{q}} }(-{\bf{q}})(0) ] \rangle
\]
\[
\sum_{ {\bf{k}}, {\bf{k}}^{'} }
\langle
 d^{\dagger}_{ (1/2){\bf{k+q/2}} }({\bf{k+q/2}})(t)
 d_{ (1/2){\bf{k-q/2}} }({\bf{k-q/2}})(t)
 d^{\dagger}_{ (1/2){\bf{k^{'}-q/2}} }({\bf{k^{'}-q/2}})(0)
 d_{ (1/2){\bf{k^{'}+q/2}} }({\bf{k^{'}+q/2}})(0)
\rangle
\]
\[
= N
(\frac{ \epsilon_{ {\bf{q}} } }{\omega_{ {\bf{q}} } })
 \mbox{ }exp(-i \mbox{ }\omega_{ {\bf{q}} }t)
\]
\[
+ \sum_{ {\bf{k}}, {\bf{k}}^{'} }
\langle B_{ {\bf{k+q/2}} }
f_{ {\bf{-k-q/2}} }e^{-i\omega_{ {\bf{k+q/2}} }t}
A_{  {\bf{k-q/2}} }f_{ {\bf{k-q/2}} }e^{-i\omega_{ {\bf{k-q/2}} }t}
A_{  {\bf{k^{'}-q/2}} }f^{\dagger}_{ {\bf{k^{'}-q/2}} }
B_{ {\bf{k^{'}+q/2}} }f^{\dagger}_{ {\bf{-k^{'}-q/2}} } \rangle
\]
\[
+ \sum_{ {\bf{k}}, {\bf{k}}^{'} }
\langle B_{ {\bf{k+q/2}} }
f_{ {\bf{-k-q/2}} }e^{-i\omega_{ {\bf{k+q/2}} }t}
B_{  {\bf{k-q/2}} }f^{\dagger}_{ {\bf{-k+q/2}} }e^{i\omega_{ {\bf{k-q/2}} }t}
B_{  {\bf{k^{'}-q/2}} }f_{ {\bf{-k^{'}+q/2}} }
B_{ {\bf{k^{'}+q/2}} }f^{\dagger}_{ {\bf{-k^{'}-q/2}} } \rangle
\]
\[
S^{>}({\bf{q}},t) = N
(\frac{ \epsilon_{ {\bf{q}} } }{\omega_{ {\bf{q}} } })
 \mbox{ }exp(-i \mbox{ }\omega_{ {\bf{q}} }t)
\]
\[
+ \sum_{ {\bf{k}} }
 exp(-i \mbox{ }(\omega_{ {\bf{k+q/2}} }+\omega_{ {\bf{k-q/2}} })t)
[ B^{2}_{ {\bf{k+q/2}} }A^{2}_{ {\bf{k-q/2}} } +
 B_{ {\bf{k+q/2}} }B_{ {\bf{-k+q/2}} }A_{ {\bf{k-q/2}} }
A_{ {\bf{k+q/2}} } ]
\]
\[
S^{<}({\bf{q}},t) =  N
(\frac{ \epsilon_{ {\bf{q}} } }{\omega_{ {\bf{q}} } })
\mbox{ }exp(i \mbox{ }\omega_{ {\bf{q}} }t)
+ \sum_{ {\bf{k}} }
 exp(i \mbox{ }(\omega_{ {\bf{k+q/2}} }+\omega_{ {\bf{k-q/2}} })t)
\]
\[
[ B^{2}_{ {\bf{k-q/2}} }A^{2}_{ {\bf{k+q/2}} } +
 B_{ {\bf{k-q/2}} }B_{ {\bf{-k-q/2}} }A_{ {\bf{k+q/2}} }
A_{ {\bf{k-q/2}} } ]
\]
and,
\[
 S^{>}({\bf{q}},t) = \langle \rho_{ {\bf{q}} }(t)  \rho_{ -{\bf{q}} }(0) \rangle
\]
\[
 S^{<}({\bf{q}},t) = \langle \rho_{ -{\bf{q}} }(0)  \rho_{ {\bf{q}} }(t) \rangle
\]
 From the above equations, it is easy to see that there is a coherent
 part corresponding to the Bogoliubov spectrum and an incoherent part which is
 due to correlations and is responsible (hopefully) for the
 roton minimum.

\end{document}